\documentclass[aps,prb,reprint,superscriptaddress]{revtex4-2}
\usepackage{hyperref}
\usepackage{graphicx}
\usepackage{color}
\usepackage{amsfonts}
\usepackage{amsmath}
\usepackage[utf8]{inputenc}
\usepackage[T1]{fontenc}
\usepackage{mathtools}
\usepackage{bbm}
\usepackage[dvipsnames]{xcolor}
\hypersetup{colorlinks=true, urlcolor=blue, citecolor = blue}

\usepackage[normalem]{ulem} 

\usepackage{braket}
\usepackage{amssymb}
\usepackage{bm}
\begin{document}

\title{Magnetic-field-free parity-protected qubit based on a narrow altermagnet Josephson junction}

\author{S. Vosoughi-nia}
\email{sakineh.vosooghi@gmail.com}
\affiliation{AGH University of Krakow, Academic Centre for Materials and Nanotechnology, al. A. Mickiewicza 30, 30-059 Krakow, Poland}

\author{M. P. Nowak}        
\email{mpnowak@agh.edu.pl}
\affiliation{AGH University of Krakow, Academic Centre for Materials and Nanotechnology, al. A. Mickiewicza 30, 30-059 Krakow, Poland}

\date{\today}

\begin{abstract}
Altermagnets provide a new route to engineer superconducting circuits without magnetic fields. We theoretically study the Andreev bound state (ABS) spectrum of a finite-width altrmagnet-based Josephson junction and show how the $d$-wave altermagnetic symmetry and geometric confinement shape its low-energy excitations. We find a clear distinction between the two $d$-wave symmetries: $d_{x^2-y^2}$ order produces spin splitting, whereas $d_{xy}$ order preserves spin degeneracy and exhibits splitting of the ABS spectrum induced by intermode hybridization. Leveraging these novel features, we propose applying a transverse electric field to tune the system and realize a magnetic-field-free, parity-protected superconducting qubit that we call \textit{altermon}.
\end{abstract}

\maketitle

\section{Introduction}
Quantum computing has emerged as a rapidly developing field, with various platforms being explored for qubit implementation, including trapped ions~\cite{PhysRevLett.74.4091, RevModPhys.75.281, PRXQuantum.4.030311, Cai_2023, Chen_2024, PhysRevA.110.L010601, 10.1063/1.5088164}, photonic systems~\cite{O_Brien_2009, 10.1063/5.0049372, Guo:2023gwg, PhysRevResearch.6.043029, doi:10.1126/science.abe8770}, spin qubits in semiconductors~\cite{PhysRevA.57.120, Kane1998ASN, RevModPhys.95.025003, Fang_2023, edlbauer2022semiconductor}, and superconducting qubits~\cite{Nielsen_Chuang_2010, arute2019quantum, clarke2008superconducting, devoret2013superconducting, PRXQuantum.6.010308, PhysRevB.106.115411, Kjaergaard2020, reed2013entanglement}. Among these platforms, superconducting qubits have attracted significant attention, both experimentally and theoretically, due to their scalability and fast operation times.
Since their inception, they have undergone significant evolution, with various designs improving coherence and control. Early implementations, such as the Cooper-pair box, relied on the charging energy of a small superconducting island connected to a reservoir via a Josephson junction (JJ)~\cite{VBouchiat_1998, nakamura1999coherent}. However, these charge qubits suffered from strong sensitivity to charge noise, leading to short coherence times. This limitation was addressed by the introduction of the transmon qubit, which operates in a regime where the Josephson energy $E_J$ is much larger than the charging energy $E_C$, thereby exponentially suppressing charge noise at the cost of reduced anharmonicity~\cite{Koch2007}. Due to its improved coherence and robust operation, the transmon has become the dominant superconducting qubit architecture, forming the basis of many modern quantum computing platforms.

Building on the transmon architecture, gatemons introduced semiconductors as tunable elements, replacing the conventional fixed Josephson junction with a semiconductor-superconductor hybrid device~\cite{Larsen2015, Casparis2018, Kringho2020}. This development allows the Josephson energy to be electrically tuned using gate voltages, providing greater flexibility and control over the qubit design. In addition, recent advances in gatemon design have focused on improving key properties such as anharmonicity and coherence times, facilitating the development of more robust and scalable qubits~\cite{deLange2015, kringhoj2020exploring, Kringhoj2020_anharmonicity, Aguado2020}.

A promising approach to achieving inherently robust superconducting qubits is the parity-protected design~\cite{smith2020superconducting}, where qubit coherence is enhanced by preserving the parity of Cooper pairs. One implementation utilizes a symmetric interferometer with two voltage-controlled semiconductor Josephson junctions, tuned into balance and threaded by a half-flux quantum, to realize a $\pi$-periodic $\cos(2\phi)$ Josephson element. This configuration significantly suppresses qubit relaxation in the protected regime~\cite{Larsen2020_parity}. Another recent approach achieved parity protection using a single highly transparent superconductor-semiconductor Josephson junction, where spin-orbit coupling and Zeeman splitting enable tunable Josephson coupling through the spin degree of freedom of Cooper pairs. This $0$-$\pi$ qubit design maintains resilience to fabrication imperfections without increasing device complexity~\cite{Guo2022_0-pi_Qubit}. Recent qubit designs explored unconventional superconductors to enhance performance. The $d$-mon qubit utilizes a planar $c$-axis junction between $d$-wave and $s$-wave superconductors for strong anharmonicity \cite{dMon2024}, while the flowermon qubit employs twisted cuprate heterostructures to engineer a novel, inherently protected qubit with high coherence \cite{Flowermon2024}.

Altermagnets, a class of collinear antiferromagnets with a momentum-dependent magnetic order parameter~\cite{PhysRevX.12.031042, PhysRevX.12.040501}, have recently attracted significant attention as a source of time-reversal symmetry breaking with zero net magnetization~\cite{Mazin2023, PhysRevB.108.184505, Fukaya_2025}. Notably, despite their vanishing magnetization, altermagnets can induce substantial spin splitting, which changes sign in different regions of the Brillouin zone. In this study, we focus on a $d$-wave magnetic order parameter, for which several potential material realizations have been proposed~\cite{PhysRevMaterials.9.024402, Jiang2025, Reichlova2024}.

Recently, the transport properties of junctions composed of altermagnets and superconductors have emerged as a topic of considerable interest~\cite{Beenakker2023, Ouassou2023, zhang2024finite, Papaj2023, Ghorashi2024, Lu2024_PRL, Cheng2024_Orientation, Giil2024_Quasiclassical, Sun2023_Andreev, alipourzadeh2025, PhysRevB.108.205410, PhysRevB.111.165406, debnath2025, PhysRevB.111.064502, PhysRevB.111.184515}. In this work, we investigate the Andreev spectrum of a finite-width altermagnetic (AM) Josephson junction, assuming that both altermagnetism and superconductivity are induced in a semiconductor (SM) nanowire via the proximity effect~\cite{kqy8-myz1}. We show that two distinct $d$-wave AM symmetries give rise to different types of Andreev energy splitting; spin splitting and splitting due to intermode (orbital) hybridization, both of which can be tuned by varying the junction parameters, paving the way for designing quantum two-level systems with highly tunable Josephson junctions~\cite{ioffe1999environmentally}. Based on this system, we propose a superconducting qubit design in which the AM material plays a crucial role. By leveraging the unique properties of the altermagnet, we demonstrate the feasibility of reaching the parity-protected regime without the need for an external magnetic field. Our approach not only advances the development of intrinsically robust qubit architectures but also opens different avenues for exploring the interplay between altermagnetism and superconductivity in quantum technologies.
\section{Model}
%
%
\begin{figure}[t]
\includegraphics[width=0.99\columnwidth, trim={63mm 248mm 65mm 24mm}, clip]{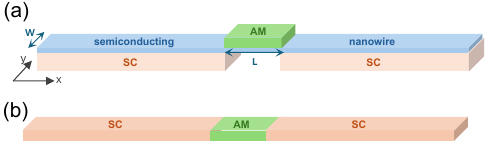}
\caption{Altermagnetic Josephson junctions of length $L$ and width $W$: (a) gatemonlike setup based on a semiconductor (SM) nanowire proximitized by superconducting (SC) and altermagnetic (AM) regions and (b) transmonlike setup based on a conventional altermagnetic Josephson junction.}\label{Fig: Schematic}
\end{figure}
%
%
We propose the schematic setup in Fig.~\ref{Fig: Schematic}(a), which is a Josephson junction consisting of a semiconducting nanowire proximitized by a pair of $s$-wave superconducting (SC) segments ($x<0$ and $x>L$) and an altermagnetic material ($0<x<L$) with $d$-wave symmetry. The length (width) of the junction is denoted by $L$ ($W$), and $L_S$ ($\gg \!\!L$) represents the length of the SC regions. We assume that superconductivity and altermagnetism are induced in the nanowire by the proximity effect.

The excitation spectrum is described by the Bogoliubov–de Gennes (BdG) Hamiltonian,
\begin{equation}\label{eq:BdG}
H(\bm{k})=\begin{pmatrix}
H_0(\bm{k}) \hspace{0.5ex}& \Delta \\[0.5ex]
\Delta^* \hspace{0.5ex}& -\sigma_yH_0^*(-\bm{k})\sigma_y
\end{pmatrix},
\end{equation}

\begin{equation}\label{eq:H0}
\begin{aligned}
H_0(\bm{k}) &= (\frac{\hbar^2 k_x^2}{2m}+\frac{\hbar^2 k_y^2}{2m}-\mu) \sigma_0\\
&+ \frac{\hbar^2}{m} t_1 k_x k_y \sigma_z + \frac{\hbar^2}{m} t_2 (k_y^2-k_x^2) \sigma_z,
\end{aligned}
\end{equation}
where $\sigma$'s are Pauli matrices in spin space, $\bm{k}=(k_x,k_y)$ is the two-dimensional electron momentum, and $\mu$ is the chemical potential which is kept the same throughout the system. $\Delta=\Delta_0e^{-i \phi}$, where $\Delta_0$ and $\phi$ are the superconducting gap and phase difference, respectively. We set $\phi=0$ at the left SC segment. The dimensionless parameters $t_1$ and $t_2$ correspond to the $d$-wave exchange interaction and depend on the altermagnet strength and the angle of the SC/AM interface relative to the crysalline axes. For a nonzero $t_1$ (and $t_2=0$), the magnetization has pure $d_{xy}$ symmetry, and for a nonzero $t_2$ (and $t_1=0$) it has pure $d_{x^2-y^2}$ symmetry. In the SC regions, we set $t_1=0=t_2$, $\Delta_0=5\times10^{-3}t$, and $L_S=2000a$. In this work, we neglect the spin-orbit coupling associated with the semiconductor nanowire in order to focus on the altermagnetism-induced parity-protection mechanism. It can be realized in semiconductor platforms based on materials with weak effective spin-orbit interaction~\cite{10.1063/1.1368156, PhysRevB.97.165401, delvecchio2026tailoringgermaniumheterostructuresquantum, wu2013mcmillanrowelloscillationslowspinorbit, 10.1063/1.117689, 10.1063/1.4941733, 10.1063/1.3377897}. The altermagnetic Josephson coupling alone is sufficient to realize a parity-protected qubit, eliminating the requirement for spin-orbit coupling. 

We note that the effective low-energy Hamiltonian in Eqs. (1) and (2) also describes the setup shown in Fig. 1(b), where the weak link is formed directly by an altermagnetic material rather than a proximitized semiconductor nanowire. This indicates that the parity-protection mechanism is not tied to the semiconductor nature of the junction but instead originates from the altermagnetic Josephson coupling. Therefore, the same mechanism can be extended to conventional altermagnetic Josephson junctions, providing a route toward transmonlike parity-protected qubits based on altermagnetic materials. The parity-protection regime can be tuned by controlling relevant junction parameters, such as the altermagnetic exchange strength or orientation, chemical potential, and junction length. However, compared with electrical tuning of the proximitized semiconductor nanowire platform, tuning these parameters may be more challenging experimentally, which motivates our focus on the gatemonlike realization shown in Fig. 1(a).

Throughout this work, we focus on the AM-proximitized semiconductor nanowire Josephson junction shown in Fig. 1(a), which provides a highly transparent, electrically tunable platform for realizing a gatemonlike parity-protected qubit. For numerical calculations, we discretize the Hamiltonian on a square tight-binding lattice using {\footnotesize KWANT}~\cite{Groth2014_Kwant}, with a lattice constant $a = 0.25$. All energies are expressed in units of $t=\hbar^2/(16ma^2)$, where we set $\hbar = m = 4a = 1$. It is important to note that the last term in the Hamiltonian, $-t_2 k_x^2$, should be treated as $-k_x t_2(x) k_x$ since $t_2$ is a position-dependent coefficient that exhibits an abrupt step at the junction interfaces~\cite{PhysRevA.52.1845}. The code used for the calculation is available in an online repository~\cite{altermon2025codes}.
\section{Results}
\subsection{Normal-region bandstructure}
%
%
\begin{figure}[b]
\includegraphics[width=0.7\columnwidth]{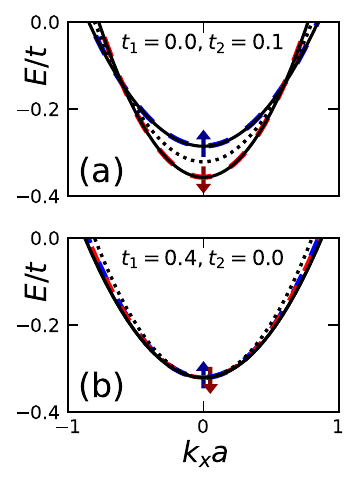}
\caption{Electronic band structure of a narrow altermagnet with a width of $W=20a$ and a chemical potential $\mu=0.5t$. (a) Pure $d_{x^2-y^2}$-wave magnetization symmetry $(t_1=0, t_2=0.1)$ and (b) pure $d_{xy}$-wave magnetization symmetry $(t_1=0.4, t_2=0)$. The dotted curves represent the band structure of the system in the absence of altermagnetism $(t_1=t_2=0)$. The blue and red dashed curves show the analytical spin-up and spin-down energy bands, respectively [see Eq.~\eqref{eq:E(k)_updown}].}\label{Fig:Lead}
\end{figure}
%
%

To study the effect of altermagnetism, we first consider a homogeneous narrow altermagnetic system in its normal state ($\Delta=0$) with $t_1=t_2=0$, assuming translational invariance [see Hamiltonian~\eqref{eq:H0}]. We then analyze how its band structure is modified by introducing the two pure magnetization terms corresponding to the $d_{x^2-y^2}$-wave ($t_1=0)$ and $d_{xy}$-wave ($t_2=0$) symmetries (see Fig.~\ref{Fig:Lead}).

As shown in Fig.~\ref{Fig:Lead}(a) and (b), in the case of pure $d_{x^2-y^2}$-wave symmetry (nonzero $t_2$), the band is split into two spin-resolved bands, whereas the spin degeneracy is preserved for the pure $d_{xy}$-wave symmetry (nonzero $t_1$). Note also that the effective mass and hence the bandwidths are modified in both cases. To better understand the impact of the $t_1$ and $t_2$ terms on the band structure, we also analytically calculated the energy bands using perturbation theory (for details, see Appendix~\ref{sec:Folding-Down method}),
\begin{equation}\label{eq:E(k)_updown}
\begin{aligned}
E_{\uparrow(\downarrow)}&(k_x)
=\frac{\hbar^2 k_x^2}{2m} \left[ 1 \mp 2t_2-\frac{2\hbar^2 t_1^2}{mE_2(1 \pm 2t_2)}(\frac{8}{3W})^2\right]\\
&- \mu + E_1(1 \pm 2t_2), 
\end{aligned}
\end{equation}
where $E_n=n^2\pi^2 \hbar^2/(2mW^2)$ with $n=1,2$ is the excitation energy in the $y$ direction. 
These analytical results, shown as blue (spin up) and red (spin down) dashed lines, are in good agreement with the numerical results (solid black lines).
\begin{figure}[b]
\includegraphics[width=0.88\columnwidth]{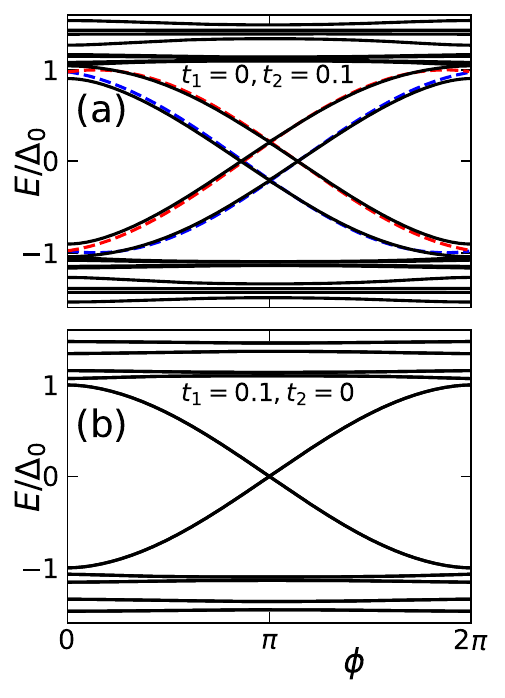}
\caption{Andreev level spectra in a single-mode ($n=1$) altermagnet Josephson junction with pure $d$-wave symmetries, $L_S=2000a$, $L=24a$, $W=20a$ and $\mu=0.5t$. Solid curves represent the numerical solution of the BdG equation on a lattice, while dashed curves illustrate the analytical model for the $t_2\neq0$ case (blue for spin up, red for spin down).}
\label{Fig:ABS_sm}
\end{figure}
\subsection{Andreev bound state spectra}
In this section, we demonstrate how altermagnetism affects Andreev levels compared to a nonmagnetic Josephson junction, for which the Andreev levels are spin degenerate. First, we consider a single-mode junction. As shown in Fig.~\ref{Fig:ABS_sm}, only the $d_{x^2-y^2}$ altermagnet ($t_2$ term) induces spin-polarized Andreev levels. This arises from spin splitting in the dispersion, as previously demonstrated in Fig.~\ref{Fig:Lead}(a). The dispersion in our system has two key features that allow us to follow the methodology developed in Ref.~\cite{PhysRevB.89.195407} to calculate Andreev levels analytically: (1) spin-polarized bands and (2) spin-dependent Fermi velocities. The latter arises from a modified effective mass for each spin flavor. In Ref.~\cite{PhysRevB.89.195407}, the Andreev levels are given by the analytical expressions for a Josephson junction based on semiconductor nanowires with spin-orbit interaction and the Zeeman effect. In our system, instead of the Zeeman energy $E_z$ we introduce the altermagnet energy $E_A$, which for pure $d_{x^2-y^2}$-wave symmetry ($t_1=0$) is given by
\begin{equation}\label{eq:E_A}
\begin{aligned}
E_A &= \left| \braket{\psi_n(y) |\frac{\hbar^2}{m}t_2(k_y^2-k_x^2)| \psi_n(y)} \right| \\
&= \left| 2t_2(2E_n-\mu) \right|,
\end{aligned}
\end{equation}
where $k_x^2=2m(\mu-E_n)/\hbar^2$. The spin-dependent Fermi velocities are given by 
\begin{equation}\label{eq:v_F}
v_{F,+(-)}\equiv v_{F\uparrow(\downarrow)}=\sqrt{\frac{2\mu-2E_n(1\pm2t_2)}{m(1\mp2t_2)}}\;.
\end{equation}
We then analytically calculate the Andreev energies for a fully transparent junction ($T=1$) and for the lowest transverse mode ($n=1$), which take the form
\begin{equation}\label{eq:E_ABS_up}
\begin{aligned}
&E_{\uparrow(\downarrow)\pm}(\phi) \\ 
& = \Delta_0 \cos{\left[ (-)\frac{\theta_A}{2} + \arccos{\left(\pm \sqrt{\frac{1+T\cos{(\phi-\phi_0)}}{2}}\right)}\right]},
\end{aligned}
\end{equation}
with
\begin{equation}\label{eq:theta_A_phi_0}
\begin{aligned}
&\theta_A = E_A L \left(\frac{1}{\hbar v_{F\uparrow}} + \frac{1}{\hbar v_{F\downarrow}} \right),\\
&\phi_0 = E_A L \left(\frac{1}{\hbar v_{F\downarrow}} - \frac{1}{\hbar v_{F\uparrow}} \right).
\end{aligned}
\end{equation}
As evident from Eq.~\eqref{eq:theta_A_phi_0}, the spin-resolved Andreev energies are sensitive to the junction length $L$. An increase in $L$ enhances the spin splitting, resulting in a greater shift of the Andreev levels. Figure~\ref{Fig:ABS_sm}(a) shows that the analytical Andreev levels (dashed lines) align well with the numerical results (solid black lines).

\begin{figure}[t]
\includegraphics[width=0.9\columnwidth]{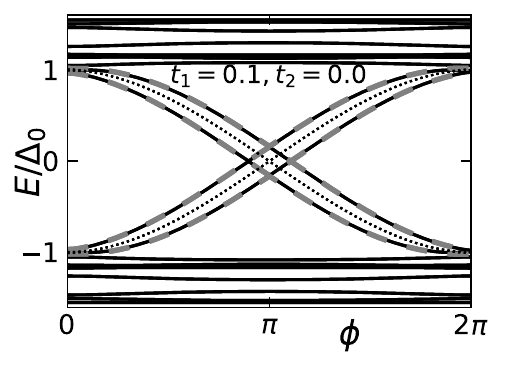}
\caption{Same as Fig.~\ref{Fig:ABS_sm}(b), but for $W=36a$ and showing twice as many Andreev levels. The dashed gray curves illustrate the two-mode ($n=2)$ BdG Hamiltonian model. Dotted black curves: nonmagnetic junction ($t_1=t_2=0$) showing spin- and mode-degenerate levels.}
\label{Fig:ABS_dm}
\end{figure}

When $t_1$ is nonzero, i.e. for $d_{xy}$ symmetry, the Andreev levels remain spin degenerate in the single-mode setup [Fig.~\ref{Fig:ABS_sm}(b)], and the spectrum is identical to that of a nonmagnetic junction ($t_1=t_2=0$). 
However, when more than one mode is considered in the junction by increasing the width, the $k_xk_y$ term couples the modes, leading to split Andreev levels. In other words, the coupling between the bands results in a shift of the Andreev levels, while their spin degeneracy is preserved. In Fig.~\ref{Fig:ABS_dm}, we show the phase dependence of the Andreev levels for $W=36a$ (solid black curves). The dotted black curves show the Andreev energies of the nonmagnetic junction ($t_1=t_2=0$), which are degenerate both in spin and in the transverse mode (band). To clarify that the shift in Andreev levels arises from band mixing rather than spin splitting, we construct a two-mode BdG Hamiltonian $H_{2m}$ in which the intermode (interband) coupling is introduced via $H_{12(21)}$ (see Appendix~\ref{sec:Folding-Down method}),
\begin{equation}
\label{H12_21}
H_{12} = -H_{21} \\
       = \frac{i\hbar^2}{m} t_1 k_x (\frac{8}{3W}) \sigma_z.
\end{equation}
The strong agreement between the solid and dashed curves validates the two-mode model and confirms that the shift is indeed caused by band mixing.
The Andreev spectrum is influenced by various parameters of the Hamiltonian and the junction~\cite{Lu2024_PRL}, including the chemical potential $\mu$ (tunable via a back gate), the AM parameters $t_1$ and $t_2$ (which depend on the angle of the AM-SC interface relative to the crystalline axes~\cite{Beenakker2023, PhysRevX.12.011028}), and the junction length $L$. Here we find that the dispersion relation of the AM part is determined by the wave-function distribution through Eqs.~\eqref{eq:H11(22)} and~\eqref{eq:H12_21}. Therefore, we propose applying a planar electric field to the AM segment along the $y$ direction, which introduces an energy term $|e|E_yy$ to the Hamiltonian (with $e=1$ for simplicity) and controls the span of wave functions across the device. In the following, we demonstrate that applying an electric field, which tunes the Andreev spectrum and consequently the Josephson potential, enables the realization of a parity-protected qubit.
\subsection{Qubit}
\begin{figure}[t]
\includegraphics[width=0.88\columnwidth]{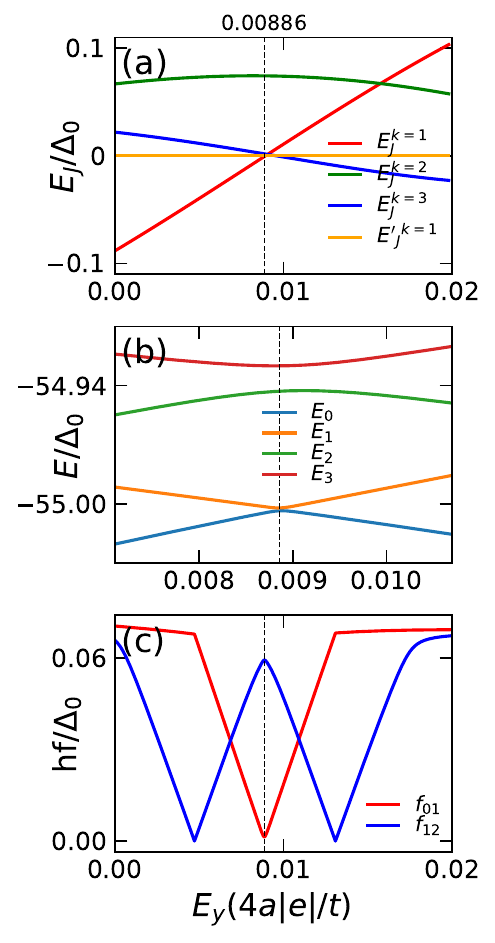}
\caption{Altermon low-energy spectrum and frequencies as a function of applied electric field $E_y$: (a) dominant components of the Josephson energy, (b) four lowest energy levels around the transition point $E_y=0.00886 \, t/(4a|e|)$, at which the two lowest levels become nearly degenerate, and (c) two lowest frequencies, $hf_{01(12)}/\Delta_0 \equiv E_{01(12)}=E_{1(2)}-E_{0(1)}$. Here we assume $E_J^{k=2}/E_C\approx20$, with $t_1=0$ and $t_2=0.3$ (pure $d_{x^2-y^2}$ altermagnetic symmetry), while all other parameters are the same as in Fig.~\ref{Fig:ABS_sm}.}
\label{Fig:t2control}
\end{figure}
\begin{figure}[t]
\includegraphics[width=0.88\columnwidth]{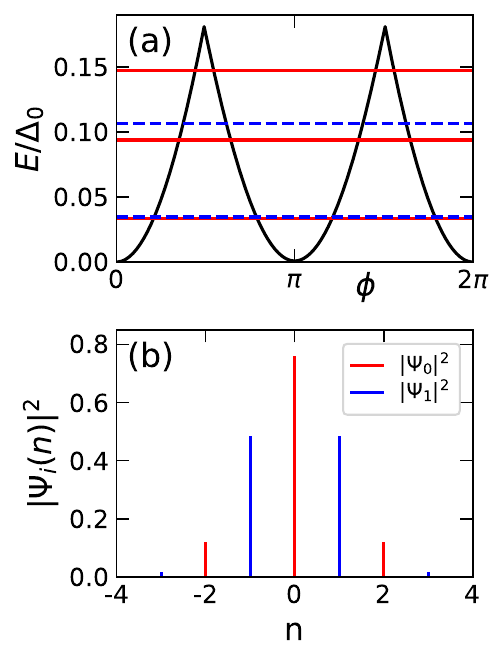}
\caption{Altermon low-energy spectrum at the transition point $E_y=0.00886\,t/(4a|e|)$. (a) Level structure and (Josephson) double-well ($\pi$-periodic) potential energy. Even and odd levels are shown by red and blue lines respectively, and the Josephson potential minima are shifted to zero energy. (b) The charge distribution of the two nearly degenerate ground states. Numerical parameters are the same as in Fig.~\ref{Fig:t2control}.}
\label{Fig:JosPot}
\end{figure}
We consider a qubit similar to the transmon (gatemon) qubit which we refer to as \textit{altermon}, and it is based on the proposed AM JJ. The altermon's Hamiltonian is thus given by
\begin{equation}\label{eq:Hq}
H_q=4E_C\hat{n}^2+U_J(\hat{\phi}).
\end{equation}
Here $\hat{n}$ is the charge operator conjugate to $\hat{\phi}$ and represents the number of Cooper pairs. $E_C$ denotes the charging energy, while $U_J(\hat{\phi})$ is the Josephson potential given in the phase basis, which we compute numerically by summing over all negative Andreev energies divided by 2~\cite{BEENAKKER1992481, PhysRevB.89.195407}. Andreev levels beyond the superconducting gap are also included in the summation because they have a (weak) phase dependence (see Figs.~\ref{Fig:ABS_sm} and~\ref{Fig:ABS_dm}) and thus contribute to the Josephson potential. To numerically calculate the altermon spectrum, we express the Hamiltonian $H_q$ in the charge basis by applying a discrete Fourier transform to $U_J(\hat{\phi})$~\cite{Larsen2020_parity, Guo2022_0-pi_Qubit},
\begin{equation}\label{eq:Hq_n}
H_q=4E_C\sum_{n} n^2 |n\rangle \langle n|+\sum_{n,k}(E_k|n\rangle \langle n+k|+\mathrm{H.c.}),
\end{equation}
where $\ket{n}$ indicates a state of $n$ Cooper pairs and $E_k=(E_J^k+iE_J'^k)/2$ denotes the complex Fourier components of the Josephson potential, representing the tunneling amplitudes (the hopping matrix element) of $k$ Cooper pairs between $\ket{n}$ and $\ket{n\pm k}$~\cite{GirvinLesHouches2014}. The real and imaginary parts of $E_k$ correspond to the cosine and sine harmonic contributions to the Josephson potential, respectively. We truncate the Hamiltonian matrix to specific $n$ and $k$ values and diagonalize it to obtain the qubit eigenstates and eigenvectors as a function of $E_y$. For a given value of $E_y$, we demonstrate that the altermon enters the parity-protected regime.
\subsubsection{Parity-protected regime for pure $d_{x^2-y^2}$ altermagnetic symmetry}
\begin{figure}[t]
\centering
\includegraphics[width=0.88\columnwidth]{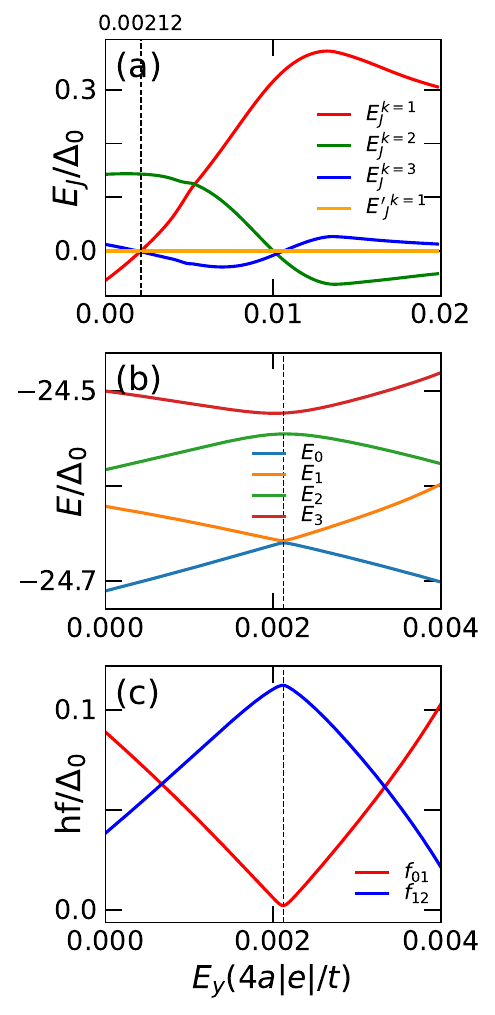}
\caption{Same as Fig.~\ref{Fig:t2control}, but for pure $d_{xy}$ altermagnetic symmetry ($t_1=0.4,t_2=0$) with a critical electric field $E_y=0.00212 \, t/(4a|e|)$, $L=W=40a$, and $\mu=0.25t$.}
\label{Fig:t1control}
\end{figure}
\begin{figure}[t]
\centering
\includegraphics[width=0.88\columnwidth]{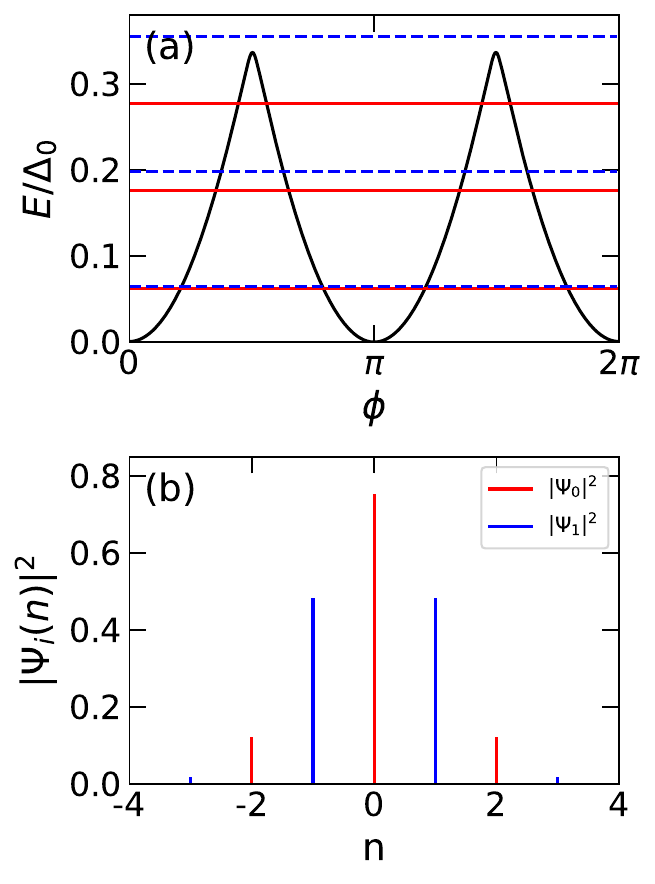}
\caption{Same as Fig.~\ref{Fig:JosPot}, but for pure $d_{xy}$ altermagnetic symmetry ($t_1=0.4,t_2=0$) at the transition point $E_y=0.00212 \, t/(4a|e|)$. Numerical parameters are the same as in Fig.~\ref{Fig:t1control}.}
\label{Fig:JosPot_t1control}
\end{figure}
%
In Figs.~\ref{Fig:t2control} and~\ref{Fig:JosPot}, we present the altermon spectrum with pure $d_{x^2-y^2}$-wave magnetization symmetry ($t_1=0$) as a function of the applied electric field $E_y$. By tuning the magnetization strength ($t_2=0.3$) and varying $E_y$, we identify a transition point at $E_y=0.00886 \, t/(4a|e|)$ (indicated by the vertical black dashed line), at which the qubit enters the parity-protected regime. In this regime, the Josephson potential transforms into a double-well ($\pi$-periodic) potential, resulting in two nearly degenerate ground states, as illustrated in Figs.~\ref{Fig:t2control}(b) and~\ref{Fig:JosPot}(a).

Figure~\ref{Fig:t2control}(a) shows the leading Fourier components of the Josephson potential as a function of $E_y$. At the transition point, the $E_J^{k=2}$ harmonic dominates, yielding a potential effectively governed by a $\cos(2\phi)$ element (a double-well structure that is $\pi$ periodic). This is a hallmark of parity-protected qubits, where single Cooper pair tunneling is suppressed and only pairs of Cooper pairs can tunnel. Consequently, the qubit exhibits two nearly degenerate ground states that differ by parity (the number of Cooper pairs) and are inherently protected against certain types of decoherence, as depicted in Fig.~\ref{Fig:JosPot}(b). Since these two states belong to opposite parity sectors, matrix elements of parity-preserving operators, such as the charge operator $\hat{n}$, are strongly suppressed, resulting in a vanishing transition matrix element $\langle\Psi_0|\hat{n}|\Psi_1\rangle$. According to Fermi's golden rule, the qubit relaxation rate is proportional to the square of the transition matrix element, $|\langle\Psi_0|\hat{n}|\Psi_1\rangle|^2$. Consequently, the suppression of the transition matrix element due to parity separation strongly suppresses relaxation between the qubit states. This symmetry-derived protection and robustness against local perturbations is a characteristic feature of parity-protected superconducting qubits with a $\pi$-periodic Josephson potential, including the proposed altermon. Our results therefore demonstrate that altermagnetic Josephson junctions constitute a new platform for realizing parity-protected superconducting qubits without requiring external magnetic fields.

In Fig.~\ref{Fig:t2control}(c), we plot the two lowest energy level differences (transition energies), which correspond to the qubit transition frequencies. At the critical electric field, a pronounced anharmonicity is observed, defined as $\alpha = E_{12}-E_{01}$, which is a key parameter in qubit design. It enables selective control of quantum transitions and suppresses leakage to higher excited states.
\subsubsection{Parity-protected regime for pure $d_{xy}$ altermagnetic symmetry}
For pure $d_{xy}$ symmetry ($t_1=0.4, t_2=0$), the altermon enters the parity-protected regime under a different set of parameters at a critical electric field $E_y=0.00212 \, t/(4a|e|)$ (see Figs.~\ref{Fig:t1control} and~\ref{Fig:JosPot_t1control}). The system exhibits the same characteristic features as discussed above for $d_{x^2-y^2}$ order---dominance of the $E_J^{k=2}$ harmonic, a $\pi$-periodic Josephson potential with nearly degenerate ground states, and strong anharmonicity in the qubit spectrum---confirming that both altermagnetic symmetries can support parity-protected qubit operation.

In realistic devices, the angle of the SC/AM interface relative to the crystalline axes is generally not restricted to the two limiting cases corresponding to the pure $d_{x^2-y^2}$ and pure $d_{xy}$ symmetries. Consequently, the altermagnetic order may slightly deviate from these pure symmetry limits or, more generally, adopt arbitrary orientations characterized by finite and comparable values of $t_1$ and $t_2$. Numerical results demonstrating the robustness of the parity-protected regime under these conditions are presented and discussed in Appendix~\ref{sec:Arbitrary_orientations}.

The robustness of the proposed altermon against disorder is another important consideration for its experimental realization. As a first step toward evaluating the impact of disorder, we investigate the effect of reduced interface transparency by introducing narrow potential barriers at the interfaces. Specifically, we model the disorder by adding local potential barriers of height $V_b$ over a narrow region of width $a$ (one lattice constant) around each interface and examine their effect on the properties of the system shown in Fig.~\ref{Fig:JosPot}.
Our numerical analysis indicates that the parity protection remains preserved for sufficiently weak interface disorder, corresponding to barrier strengths $V_b < 0.1\Delta_0$. In the range $0.1\Delta_0 \lesssim V_b \lesssim 0.2\Delta_0$, the $\pi$-periodic double-well Josephson potential and the qubit spectrum (including the nearly degenerate ground-state doublet) remain essentially unchanged, with only negligible parity leakage in the charge distribution of corresponding disordered qubit states. For stronger barriers (e.g. $V_b=\Delta_0$), the parity-protected regime is destroyed and the $\pi$ periodicity is removed, although the double-well structure itself remains largely unchanged. These findings are encouraging in view of experimental realizations, as semiconductor nanowire Josephson junctions are routinely fabricated with high transparency. Therefore, we expect the parity-protected altermon to remain robust against sufficiently weak disorder and interface imperfections. A comprehensive study of the influence of different disorder mechanisms on the altermon is beyond the scope of the present work but constitutes an interesting direction for future research.
\section{Summary and conclusions}
In this work, we explored the interplay between altermagnetism and superconductivity in a finite-width Josephson junction and its implications for qubit design. We showed that the Andreev bound state (ABS) spectrum is highly sensitive to the symmetry of the altermagnetic order parameter. For $d_{x^2-y^2}$-type altermagnetism, we observe a clear spin splitting in both the normal-state band structure and the ABS spectrum, resulting in a spin-polarized ABS spectrum that can be tuned via junction parameters. This spin splitting remains robust in the presence of multiple transverse modes. In contrast, $d_{xy}$-type order preserves spin degeneracy but leads to a splitting of the ABS spectrum through intermode hybridization in the multimode regime. For clarity, we illustrated this effect explicitly for the two-mode case. These findings highlight two qualitatively distinct mechanisms by which altermagnetism can shape the ABS spectrum in Josephson junctions. Building on this understanding, we proposed the altermon, a superconducting qubit that can be electrically tuned into a parity-protected regime without requiring an external magnetic field. These results highlight AM Josephson junctions as a promising platform for electrically controllable, intrinsically robust superconducting qubits and, more broadly, for exploring the interplay between altermagnetism and superconductivity in quantum devices.

We note in closing that the recent realization of a semiconducting $d$-wave altermagnet~\cite{zhang2026keyrolechargedisproportionation} may provide an even simpler platform for implementing the electrically tunable altermon by eliminating the need to induce altermagnetic order in a semiconducor nanowire via the proximity effect.
\section*{Acknowledgments}
This work was supported by the National Science Center, Poland (NCN) Agreement No. UMO-2020/38/E/ST3/00418. We gratefully acknowledge the Polish high-performance computing infrastructure PLGrid (HPC Center: ACK Cyfronet AGH) for providing computer facilities and support within computational grant no. PLG/2025/018486. We thank J. H. Correa for drawing our attention to altermagnetic materials. S.V. thanks C.W.J. Beenakker and Tomohiro Yokoyama for helpful discussions.
\appendix
\numberwithin{equation}{section}
\renewcommand{\thesubsection}{\arabic{subsection}}
\section{Derivation of the analytical formula for the band structure of the Altermagnet}\label{sec:Folding-Down method}
In this appendix, we provide a detailed demonstration of the down-folding method, which reduces a two-mode Hamiltonian to a single mode, simplifying it for analytical diagonalization.
We begin with the Hamiltonian~\eqref{eq:H0} from the main text, which corresponds to an infinite quantum well of width $W$ ($0<y<W$), with eigenfunctions $\psi_n(y)$ and excitation energies $E_n$ given by 
\begin{equation}\label{Psi-En}
\psi_n(y) = \sqrt{\frac{2}{W}} \sin(\frac{n \pi y}{W}), \qquad E_n = \frac{n^2 \pi^2 \hbar^2}{2mW^2}.    
\end{equation}

In the low-energy approximation, the system can be described in the framework of the two-band Hamiltonian. For this purpose, we write the Hamiltonian in the basis of the two lowest eigenstates, and we obtain
\begin{equation}\label{eq:2band_H}
H_{2\text{m}} = \begin{pmatrix}
H_{11} \hspace{0.5ex}& H_{12} \\[0.5ex]
H_{21} \hspace{0.5ex}& H_{22}
\end{pmatrix}, 
\end{equation}
with the diagonal and off-diagonal elements given by
\begin{multline}\label{eq:H11(22)}
H_{11(22)} = \braket{\psi_{1(2)} | H_0 | \psi_{1(2)}} \\
= (\frac{\hbar^2 k_x^2}{2m}+E_{1(2)}-\mu) \sigma_0 - (\frac{\hbar^2}{m} k_x^2 - 2E_{1(2)})t_2 \sigma_z,
\end{multline}
and 
\begin{equation}
\label{eq:H12_21}
H_{12} = \braket{\psi_1 | H_0 | \psi_2} = \frac{i\hbar^2}{m} t_1 k_x (\frac{8}{3W}) \sigma_z = -H_{21}.
\end{equation}
where $E_n= \braket{\psi_n | k_y^2 | \psi_n}\hbar^2/2m$ and $k_y=n\pi/W$.
Note that the $t_2$ term affects only the diagonal components of the Hamiltonian~\eqref{eq:2band_H}, implying that its impact on the band structure can be calculated independently for each transverse mode. In contrast, the $t_1$ term appears only in the off-diagonal components, which means it mixes the transverse modes.

Using the down-folding method, 
\begin{equation}
\label{eq:H(E)}
\mathcal{H}(E)=H_{11}-H_{12}(H_{22}-E)^{-1}H_{21},
\end{equation}
the $4\times4$ Hamiltonian~\eqref{eq:2band_H} can be reduced to a $2\times2$ effective Hamiltonian. Based on the fact that $E_2$ is the largest energy in the system, the expression $(H_{22}-E)^{-1}$ which is a diagonal matrix, can be simplified by expanding each of its elements into the Taylor series around $E_2$ and limiting only to the first term,
\begin{multline}\label{Taylor}
\frac{1}{\frac{\hbar^2 k_x^2}{2m}(1-2t_2)-\mu+E_2(1+2t_2)-E}= \frac{1}{E_2(1+2t_2)}\\
-\frac{1}{[E_2(1+2t_2)]^2}\left(\frac{\hbar^2 k_x^2}{2m}(1-2t_2)-\mu-E\right)+\dots\, , 
\end{multline}
and therefore,
\begin{equation}
\label{eq:H22-E}
(H_{22}-E)^{-1} \simeq \begin{pmatrix}
     \frac{1}{E_2(1+2t_2)} & 0\\
     0 & \frac{1}{E_2(1-2t_2)}
 \end{pmatrix}.
 \end{equation}
 The down-folding method leads to 
 \begin{equation}
\label{eq:H(E)_final}
\begin{aligned}
 \mathcal{H}(E)&=H_{11}-H_{12}(H_{22}-E)^{-1}H_{21}\\
 &=\begin{pmatrix}
     E_\uparrow(k_x) & 0\\[0.5ex]
     0 & E_\downarrow(k_x)
 \end{pmatrix},   
\end{aligned}
\end{equation}
where
\begin{widetext}
\begin{equation}\label{eq:E(k)_up}
 E_\uparrow(k_x)=\frac{\hbar^2 k_x^2}{2m} \left[ 1-2t_2-\frac{2\hbar^2 t_1^2}{mE_2(1+2t_2)}(\frac{8}{3W})^2\right]- \mu + E_1(1+2t_2), 
\end{equation}
and
\begin{equation}\label{eq:E(k)_down}
E_\downarrow(k_x)=\frac{\hbar^2 k_x^2}{2m} \left[ 1+2t_2-\frac{2\hbar^2 t_1^2}{mE_2(1-2t_2)}(\frac{8}{3W})^2\right]- \mu + E_1(1-2t_2) 
\end{equation}
\end{widetext}
are the spin-resolved dispersions. As we saw in Fig.~\ref{Fig:Lead}, this approximation works well and the calculated dispersions $E(k)$ are in good agreement with the numerical tight-binding results.
%
\renewcommand{\thesubsection}{\arabic{subsection}}
\section{Parity-protected regime beyond pure $d_{x^2-y^2}$- and $d_{xy}$-wave altermagnetic symmetries  ($t_1,t_2\neq0$)}\label{sec:Arbitrary_orientations}
\setcounter{figure}{0}
\renewcommand{\thefigure}{B\arabic{figure}}
In the main text, we focused on the two pure $d_{x^2-y^2}$-wave and $d_{xy}$-wave altermagnetic symmetries. In realistic experimental devices, however, a possible relative misalignment between the superconductor and the altermagnet may result in an altermagnetic order that is neither purely $d_{x^2-y^2}$ nor purely $d_{xy}$. To investigate the robustness of the proposed altermon against such small deviations, we reconsider the parity-protected regime shown in Fig.~\ref{Fig:JosPot} of the main text, which corresponds to a pure $d_{x^2-y^2}$-wave altermagneic symmetry ($t_1=0,t_2=0.3$), and gradually introduce a finite $d_{xy}$ component by allowing $t_1\neq0$ while keeping $t_2=0.3$.

\begin{figure}[t]
\centering
\includegraphics[width=0.88\columnwidth]{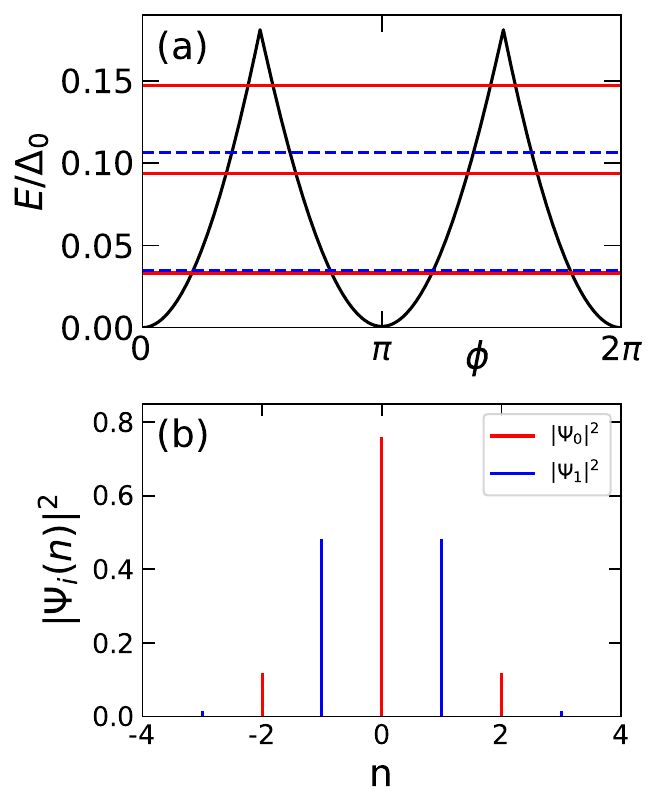}
\caption{Same as Fig.~\ref{Fig:JosPot} of the main text, but for a small deviation from pure $d_{x^2-y^2}$-wave altermagnetic symmetry; $t_1=0.005$, and $t_2=0.3$.}
\label{Fig:orientation005_JosPot_t2control}
\end{figure}

\begin{figure}[t]
\centering
\includegraphics[width=0.88\columnwidth]{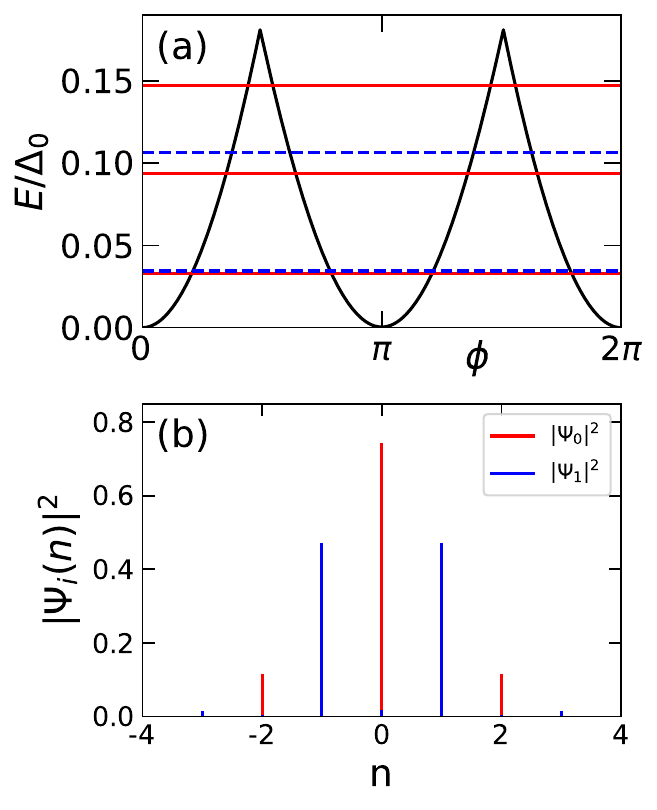}
\caption{Same as Fig.~\ref{Fig:JosPot} of the main text, but for a small deviation from pure $d_{x^2-y^2}$-wave altermagnetic symmetry; $t_1=0.015$, and $t_2=0.3$.}
\label{Fig:orientation015_JosPot_t2control}
\end{figure}

Our numerical results demonstrate that the parity-protected regime remains perfectly preserved for $0\leq t_1<0.006$ (see, for example, Fig.~\ref{Fig:orientation005_JosPot_t2control} for $t_1=0.005$). For larger values, $0.006\lesssim t_1<0.02$, only negligible parity leakage is observed, while the $\pi$-periodic double-well Josephson potential and the qubit spectrum remain essentially unchanged (see, for example, Fig.~\ref{Fig:orientation015_JosPot_t2control} for $t_1=0.015$). These results indicate that the altermon is robust against small deviations from the pure altermagnetic symmetries arising, for example, from a relative misalignment between the superconductor and the altermagnet. 

\begin{figure}[b]
\centering
\includegraphics[width=0.88\columnwidth]{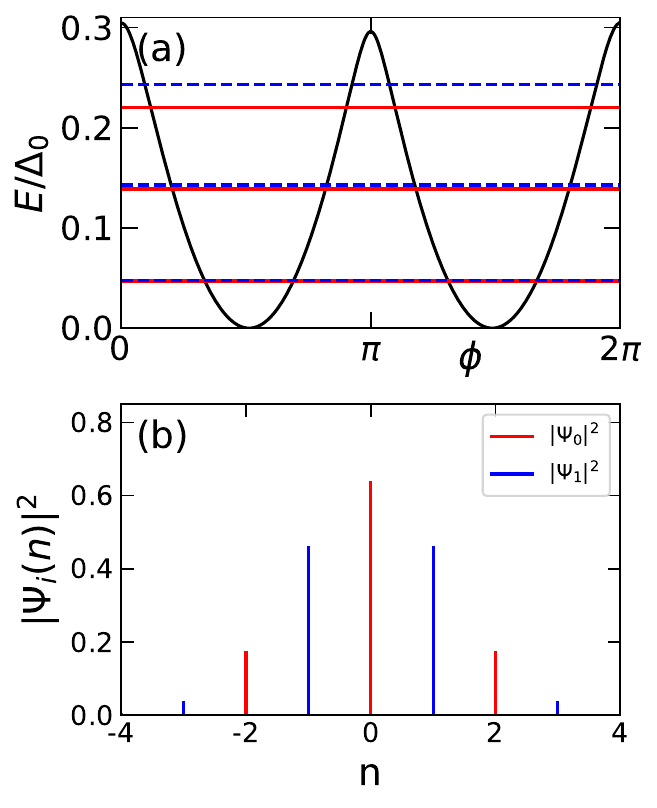}
\caption{Same as Fig.~\ref{Fig:JosPot} of the main text ($t_2=0.3$), but for $t_1=0.1$, $L=20a$, and $W=36a$ at the transition point $E_y=0.01\,t/(4a|e|)$.}
\label{Fig:phi_protection}
\end{figure}

To further demonstrate that the parity-protected regime is not restricted to the vicinity of the pure symmetry limits, we also consider a representative configuration away from these limiting cases, with finite and comparable values of the altermagnetic parameters, $t_1=0.1$ and $t_2=0.3$. The resulting Josephson potential is shown in Fig.~\ref{Fig:phi_protection}. Although the potential minima are no longer located at $0$ and $\pi$, the parity-protection characteristics are preserved. We expect that the $0$-$\pi$ parity-protected regime is also accessible after further tuning of the parameters. This example demonstrates that the proposed altermon is not restricted to the pure $d_{x^2-y^2}$- or $d_{xy}$-wave symmetry limits and can also be realized for arbitrary orientations of the altermagnetic order.

\clearpage


%


\end{document}